\documentclass{elsart}

\usepackage{amsmath,amssymb,latexsym,url,graphicx}

\usepackage{graphics}
\usepackage{psfig}
\usepackage{color}

\def\deg{{\rm deg}}

\def\qed{\hfill  \framebox(5,5){}}

\def\Res{{\rm Res}}

\newtheorem{theorem}{{\bf Theorem}}
\newtheorem{remark}{{\bf Remark}}
\newtheorem{definition}[theorem]{{\bf Definition}}
\newtheorem{corollary}[theorem]{{\bf Corollary}}
\newtheorem{proposition}[theorem]{{\bf Proposition}}
\newtheorem{lemma}[theorem]{{\bf Lemma}}
\newtheorem{example}{{\bf Example}}

\begin{document}
\begin{frontmatter}



\title{Analyzing the Topology Types arising in a Family of Algebraic Curves Depending
On Two Parameters}

\author{Juan Gerardo Alcazar\thanksref{proy}},
\ead{juange.alcazar@uah.es}

\address{Departamento de Matem\'aticas, Universidad de Alcal\'a,
E-28871-Madrid, Spain}

\thanks[proy]{Author supported by the Spanish `` Ministerio de
Educaci\'on y Ciencia" under the Project MTM2005-08690-C02-01.}

\begin{abstract}
Given the implicit equation $F(x,y,t,s)$ of a family of algebraic
plane curves depending on the parameters $t,s$, we provide an
algorithm for studying the topology types arising in the family.
For this purpose, the algorithm computes a finite partition of the
parameter space so that the topology type of the family stays
invariant over each element of the partition. The ideas contained
in the paper can be seen as a generalization of the ideas in
\cite{JGRS}, where the problem is solved for families of algebraic
curves depending on one parameter, to the two-parameters case.
\end{abstract}
\end{frontmatter}

\section{Introduction}

The computation of the topology of algebraic sets is an active
research topic. In this sense, the topology of plane algebraic
curves has been extensively addressed (see \cite{Arnon},
\cite{GianniTraverso}, \cite{Lalo}, \cite{Hong} and many others).
More recently, the computation of the topology of space algebraic
curves (see \cite{JG-Sendra}, \cite{Elka}, \cite{Gatellier},
\cite{Niang}) and of algebraic surfaces (see \cite {Cheng},
\cite{FortunaJSC}, \cite{Gianni}) has also been studied;
furthermore, related to the topological study of surfaces, the
problem of determining the topology types arising in the family of
level curves of an algebraic surface has been considered by some
authors (see \cite{JGRS}, \cite{EACA}, \cite{Mourrain}) under
different perspectives. Clearly, this last problem is analogous to
the question of determining the topology types appearing in a
family of plane algebraic curves depending on a parameter. In
addition, in \cite{Lazard} the authors solve the problem of
determining the solutions of a zero-dimensional polynomial system
depending on several parameters; this problem can be interpreted
as the computation of the topology types of a zero-dimensional
variety, depending on several parameters.

In this paper we address the problem of studying the topology
types arising in a family of plane algebraic curves depending
algebraically on two parameters. From Hardt's Semialgebraic
Triviality Theorem (see Theorem 5.46 in \cite{Basu}), it is known
that the number of topology types arising in such a family is
finite. In order to determine them, here we provide an algorithm
that determines a partition of the parameter space so that the
topology type of the family is constant over each element of the
partition. The algorithm generalizes the ideas in \cite{JGRS}
(where families of algebraic curves depending on one parameter are
considered) to the two-parameters case.

More precisely, in \cite{JGRS} it is proven that, given the
implicit equation $F(x,y,t)$ of a family of algebraic curves
depending on the parameter $t$, the parameter values where the
topology type of the family may change are contained in the set of
real roots of a double discriminant $R(t)$ of $F$; so, in between
two consecutive real roots of this polynomial, the topology type
of the family stays the same (see Subsection \ref{top-one-param}
for more details). Thus, a finite partition of the parameter space
(${\Bbb R}$, in this case) with the property that the topology
type stays invariant over each element of the partition, is
derived. When the family depends not on one, but on two
parameters, the double discriminant $R$ is a polynomial in two
variables (the parameters), and therefore it defines an algebraic
variety over ${\Bbb R}^2$. Hence, the geometry of this variety has
to be analyzed in order to compute a decomposition of the
parameter space (in our case, ${\Bbb R}^2$) with similar
properties. In this sense, the main result of this work is an
algorithm for computing a partition of ${\Bbb R}^2$ into cells of
dimensions $0,1,2$ so that the topology type of the family stays
invariant along each cell. For this purpose, the tools that we use
are, essentially, McCallum's notion of delineability (see
\cite{Arnon-Collins}), properties of resultants and its
specialization, and properties of analytic functions in several
variables and {\it germs}, basically taken from \cite{Abhyankar} and \cite{Jong}.

The structure of this paper is the following. In Section
\ref{prelim}, we review the main ideas for the one-parameter case,
we recall the notion of delineability and some related results,
and we introduce some notation and hypotheses for the
two-parameters case. In Section \ref{general-case} we thoroughly
analyze the two-parameters case and we give a full algorithm. In
Section \ref{sec-generalization} we present some examples
illustrating the algorithm.

\section{Preliminaries} \label{prelim}


\subsection{Topology of families of algebraic curves depending on
a parameter} \label{top-one-param}

Let $F\in {\Bbb R}[x,y,t]$ be a polynomial not containing any
factor just depending on the variable $t$. Thus, $F(x,y,t)=0$
defines a family of plane algebraic curves depending on the
parameter $t$, i.e. for all $t_0\in {\Bbb R}$, $F(x,y,t_0)=0$
defines an affine plane algebraic curve $F_{t_0}$. We say that two
members $F_{t_0}$, $F_{t_1}$ of the family have the same {\it
topology type}, if there exists an homeomorphism of the plane into
itself transforming $F_{t_0}$ into $F_{t_1}$; in that case, it
follows that the curves defined by $F_{t_0}$, $F_{t_1}$ have the
same shape. Then, one may address the problem of determining the
topology types arising in the family. In \cite{JGRS}, an analogous
problem, namely the computation of the topology types arising in
the family of level curves to a given algebraic surface, is
considered. So, in the sequel we will recall the main ideas in
\cite{JGRS}.

First of all, we assume that the following hypotheses on the
family $F$ hold: (i) $F$ contains no factor only depending on the
variable $t$; (ii) $F$ is square-free; (iii) the leading
coefficient of $F$ w.r.t. the variable $y$ does not depend on the
variable $x$. Note that (iii) can always be achieved by applying
if necessary a change of coordinates of the type
${x=aX+bY,y=cX+dY}$, which does not change the topology of the
family. Furthermore, we consider the following definition:

\begin{definition} \label{def-critical-set}
Let ${\mathcal C}$ be a finite subset ${\mathcal C}\subset {\Bbb
R}$, and let ${\mathcal C}=\{a_1,\ldots,a_r\}$ where
$a_1<\cdots<a_r$. Moreover, let $a_0=-\infty$, $a_{r+1}=\infty$.
We say that ${\mathcal C}$ is a {\sf critical set} of the family
defined by $F$, if given $t_i,t_{i+1}\in {\Bbb R}$ verifying that
$[t_i,t_{i+1}]\cap {\mathcal C}=\emptyset$, the topology types of
$F_{t_i}$ and $F_{t_{i+1}}$ are equal.
\end{definition}

In other words, a critical set ${\mathcal C}$ of a family is a
finite real set containing all the parameter values where the
topology type may change.

Now let us introduce the following two polynomials; here,
$D_w(G):=\Res_w(G,\frac{\partial G}{\partial w})$, and $\sqrt{G}$
denotes the square-free part of $G$. Also, abusing of language, in
the sequel we will refer to $D_w(G)$ as the ``discriminant" of $G$
w.r.t. the variable $w$ (notice that usually the discriminant
denotes the result of dividing out the resultant $D_w(G)$ by the
leading coefficient). Then we define
\[ M(x,t):=\sqrt{D_y(F(x,y,t))}, \mbox{
}R(t):=D_x(M(x,t))
\]Furthermore, $M(x,t):=0$ when $\deg_y(F)=0$, and $R(t):=0$ when $\deg_x(M)=0$.
Then the following theorem holds (see \cite{JGRS} for a proof of
this result).


\begin{theorem} \label{main-theorem-top}
Let $F$ satisfy the preceding hypotheses. Then the following
statements hold:
\begin{itemize}
\item [(1)] If $R$ is not identically zero, then the set
of real roots of $R$, is a critical set of $F$. If $R$ has no real
roots, then the elements of the family show just one topology
type.
\item [(2)] If $R$ is identically zero, then there are two
possibilities:
\begin{itemize}
\item [(i)] $M=0$, in which case
$F=F(x,t)$; here, the set of real roots of $D_x(F)$ is a critical
set.
\item [(ii)] $M\neq 0$, but $M=M(t)$; here, the set of real roots
of $M$ is a critical set.
\end{itemize}
\end{itemize}
\end{theorem}

The elements $a_i$ of a critical set ${\mathcal C}$ induce a
finite partition of the parameter space (${\Bbb R}$, in this
case). The elements of this partition are, on one hand, the $a_i$,
and on the other hand, the intervals $(a_i,a_{i+1})$. So, each
element of the partition gives rise to one topology type. In order
to describe these topology types, it suffices to consider one
$t$-value for each element of the partition; then, the topologies
of the corresponding curves can be described, for instance,  by
using the algorithm in \cite{Lalo}, \cite{Hong}.


Furthermore, if $D_y(F)$ is square-free (which typically happens
when $F$ is non-sparse) then $R$ is an iterated discriminant;
then, results on the structure of iterated discriminants (see
\cite{AMS-Mourrain} and \cite{MEGA-Lazard}) can be applied in
order to efficiently compute the real roots of $R$.

Moreover, by using basic properties of resultants one may easily
see that the following lemma, that will be used later on the
paper, holds. This result provides a geometrical interpretation of
the polynomial $M$ defined above. Here, we use the following
definition of regular, critical and singular point of a plane
algebraic curve; namely, given a polynomial $g\in {\Bbb R}[x,y]$
and a point $P$ verifying that $g(P)=0$, we say that it is: (i)
{\it regular}, if $\frac{\partial g}{\partial y}(P)\neq 0$; (ii)
{\it critical}, if $\frac{\partial g}{\partial y}(P)=0$; (iii)
{\it singular}, if $\frac{\partial g}{\partial
x}(P)=\frac{\partial g}{\partial y}(P)=0$.

\begin{lemma}\label{M-crit-points}
If $(\bar{x},\bar{y})\in {\Bbb C}^2$ is a critical point of the
curve defined by $F(x,y,\bar{t})$, then $M(\bar{x},\bar{t})=0$.
\end{lemma}


Furthermore, Lemma \ref{M-crit-points} provides the following
corollary. Here, we consider the algebraic surface $S$ defined by
the polynomial $F(x,y,t)$ in the Euclidean space with coordinates
$\{x,y,t\}$. This result will be useful in the next section.

\begin{corollary} \label{singular-surface}
Assume that $R\neq 0$, and let ${\mathcal M}$ be the curve defined
by the polynomial $M$ on the $xt$-plane. Moreover, let ${\mathcal
C}$ be the set of real roots of $R$, and let $I\subset {\Bbb R}$
so that $I\cap {\mathcal C}=\emptyset$. Then every singular point
of $S$ with $t\in I$ projects onto the $xt$-plane as a point of
${\mathcal M}$.
\end{corollary}

\subsection{Preliminaries on delineability} \label{prelim-delineability}

In \cite{JGRS}, we used as a fundamental tool the notion of {\it
delineability}; in this paper, we will also make use of this
notion and of some related results, proven in \cite{JGRS}, that we
summarize in this subsection.

Essentially, a square-free polynomial $F\in {\Bbb
R}[x_1,\ldots,x_n]$ is said to be {\sf delineable} over a manifold
${\mathcal T}\subset {\Bbb R}[x_1,\ldots,x_{n-1}]$ (for example,
an open subset), if the zero set of $F$ over ${\mathcal T}$ is the
disjoint union of the graphs of several analytic functions
$W_1,\ldots,W_r$, where $W_i:{\mathcal T} \longmapsto {\Bbb R}$. A
more detailed definition, taken from \cite{Arnon-Collins}, is
given now.

\begin{definition} \label{def-delineab}
Let $\breve{x}$ denote the $(n-1)$-tuple $(x_1,\ldots,x_{n-1})$.
An $n$-variate polynomial $F(\breve{x},x_n)$ over the reals is
said to be {\sf (analytic) delineable} on a submanifold ${\mathcal
T}$ of ${\Bbb R}^{n-1}$, if it holds that:
\begin{itemize}
\item [1.] the portion of the real variety of $F$ that lies in the
cylinder ${\mathcal T} \times {\Bbb R}$ over ${\mathcal T}$
consists of the union of the function graphs of some $r\geq 0$
analytic functions $W_1<\cdots<W_r$ from ${\mathcal T}$ into
${\Bbb R}$.
\item [2.] there exist positive integers $m_1,\ldots,m_r$ such
that for every $a\in {\mathcal T}$, the multiplicity of the root
$W_i(a)$ of $F(a,x_n)$ (considered as a polynomial in $x_n$ alone)
is $m_i$.
\end{itemize}
Furthermore, the $W_i$ in the condition 1 of the definition above
are called {\sf real roots} of $F$ over ${\mathcal T}$.
\end{definition}

Moreover, in \cite{Arnon-Collins} a sufficient condition for a
polynomial to be delineable is provided (see pp. 246 in
\cite{Arnon-Collins}). This condition is used in \cite{JGRS} in
order to prove the following result (see Section 4 of \cite{JGRS}
for a proof of the statements in this lemma); here, we recall the
definitions of the polynomials $M=M(x,t)$ and $R=R(t)$ stated in
the preceding subsection.

\begin{lemma} \label{real-roots}
Assume that $R$ is not identically $0$, and let $a_1<\cdots<a_r$
be the real roots of $R$. Then the following statements are true:
\begin{itemize}
\item [(1)] The polynomial $M$ is delineable over each
$(a_j,a_{j+1})$. The real roots of $M$ over each interval
$(a_j,a_{j+1})$ are denoted as $X_k$ and the graphs of the $X_k$
are denoted as ${\mathcal X}_k$.
\item [(2)] The polynomial $F$ is delineable over each ${\mathcal
X}_k$. The real roots of $F$ over ${\mathcal X}_k$, are denoted as
$Y_l$; the graphs of the $Y_l$'s are denoted as ${\mathcal Y}_l$.
\item [(3)] The polynomial $F$ is delineable over each open
region \[C_{j,k}=\{(x,t)\in {\Bbb R}^2|\mbox{ }t\in
(a_j,a_{j+1}),X_k(t)<x<X_{k+1}(t)\},\]The real roots of $F$ over
these regions are denoted as $V_i$; the graphs of the $V_i$'s are
denoted as ${\mathcal V}_i$.
\end{itemize}
\end{lemma}

Similar results hold when $R=0$. In Figure 1, you may see the
geometrical meaning of the functions $X_k, Y_l, V_i$ in the
statement of Lemma \ref{real-roots}. In this picture it is
implicitly assumed that the ${\mathcal Y}_l$'s and the ${\mathcal
V}_i$'s join properly, in the sense that the topological closure
of a ${\mathcal V}_i$ contains just one ${\mathcal Y}_l$. This
result is rigorously proven in \cite{JGRS} (see Lemma 11 in
\cite{JGRS}). A straightforward consequence of this fact and of
delineability properties is the following lemma, which will be
important for our purposes.

\begin{lemma} \label{main-point}
Assume that $R\neq 0$, and let $a_1<\cdots<a_r$ be the real roots
of $R$. Then along each interval $(a_j,a_{j+1})$, the relative
positions of the ${\mathcal V}_i$'s, the ${\mathcal Y}_l$'s, and
of the ${\mathcal V}_i$'s w.r.t. the ${\mathcal Y}_l$'s, stay
invariant.
\end{lemma}

\begin{figure}[ht]
\begin{center}
\centerline{
\psfig{figure=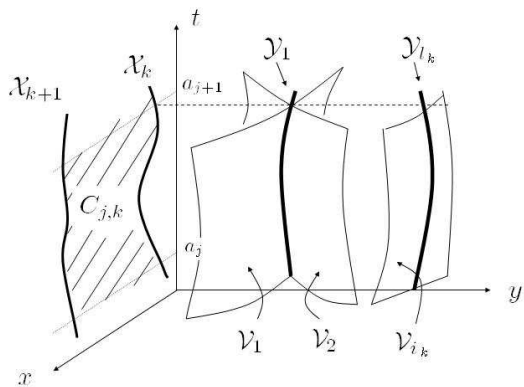,width=8cm,height=6cm}}
\end{center}
\caption{Functions in Lemma \ref{real-roots}}
\end{figure}

A similar result holds when $R=0$. Finally, we recall from
\cite{McCallum} the following result, which will be used
later (see Theorem 2.2.3 and Theorem 2.2.4 in \cite{McCallum}
for a proof)

\begin{lemma} \label{vis-connected}
The ${\mathcal X}_k$'s, the ${\mathcal Y}_l$'s and the ${\mathcal
V}_i$'s are connected sets.
\end{lemma}

\subsection{Hypotheses and notation.}
\label{hypot}

In our paper we consider the two-parameters case (see Section
\ref{general-case}). So, in this subsection we introduce the
required hypotheses and notation for this case. Thus, in the
sequel we assume to be working with a square-free polynomial $F\in
{\Bbb R}[x,y,t,s]$, containing no factor only depending on the
parameters $t,s$, and where the leading coefficient of $F$ w.r.t.
the variable $y$ does not depend on the variable $x$. As in the
one-parameter case, this last condition can always be achieved by
applying if necessary an affine transformation involving only
$x,y$. Also, in the sequel we will denote the substitution
$F(x,y,t_i,s_i)$, as $F_{t_i,s_i}$. Analogously, $F_{t_i}$ and
$F_{s_i}$ will denote the substitutions $F(x,y,t_i,s)$,
$F(x,y,t,s_i)$, respectively; note that, since by assumption $F$
contains no factor only depending on $t,s$, $F_{t_i}$ and
$F_{s_i}$ cannot be the zero polynomial. Hence, $F_{t_i}$ (resp.
$F_{s_i}$) defines a family of algebraic curves depending on the
parameter $s$ (resp. $t$); therefore, according to the notation
introduced in Subsection 2.1, for these families we would obtain
polynomials $M_{t_i},R_{t_i}$ (resp. $M_{s_i},R_{s_i}$)
 so that
the set of real roots of $R_{t_i}$ (resp. $R_{s_i}$) would be a
critical set of $F_{t_i}$ (resp. $F_{s_i}$).

In addition, as in the one-parameter case, we define the following
two polynomials:
\[M(x,t,s):=\sqrt{D_y(F(x,y,t,s))},\mbox{ }R(t,s):=\
D_x(M(x,t,s))\]Furthermore, $M(x,t):=0$ when $\deg_y(F)=0$, and
$R(t,s):=0$ when $\deg_x(M)=0$. The relationship between the
specialization of these polynomials in $t_i$ (resp. $s_i$) and the
polynomials $M_{t_i},R_{t_i}$ (resp. $M_{s_i},R_{s_i}$) defined
before, is analyzed in Subsection \ref{specialization}.

Also, whenever the polynomial $R=R(t,s)$ is not identically 0, by
applying if necessary a linear change of coordinates just
involving $x,y$ one may also assume that the leading coefficient
with respect to $y$ of the resultant $\Res_s(F,R)$ does not depend
on $x$. This assumption will be needed in Subsection
\ref{delicate-part}.

\section{The two-parameters case} \label{general-case}

Here, we consider the problem of analyzing the topology types
arising in a family of algebraic curves depending on two
parameters. Thus, along this section we assume that we are working
with a family defined by $F(x,y,t,s)$, where $t,s$ are parameters,
and $F$ satisfies the hypotheses made explicit in Subsection
\ref{hypot}. In order to solve our problem, first we will focus on
the case when the polynomial $R(t,s)$ defined in Subsection
\ref{hypot} is not identically zero; the special case when it is
the zero polynomial will be treated at the end of the section.
Under this assumption, the curve ${\mathcal R}$ defined by the
polynomial $R(t,s)$ divides the real plane, with coordinates
$t,s$, into finitely many open regions, namely the connected
components of ${\Bbb R}^2 \backslash {\mathcal R}$ (for example,
in Fig. 2 the curve ${\mathcal R}$ plotted there divides the plane
into three open regions). Notice that, by computing a C.A.D. of
${\mathcal R}$, these regions correspond to the union of finitely
many 2-dimensional cells which can be described from the C.A.D.
Thus, we will prove that:
\begin{itemize}
\item [(i)] The topology type of the family stays the same
along each of these open regions (see
 Theorem \ref{top-open-regions} in Subsection \ref{behav}).
\item [(ii)] For the remaining topology types, i.e. the topology types
over ${\mathcal R}$, one can compute a partition of this curve
into finitely many 1-dimensional and 0-dimensional cells, so that
the topology type stays the same over each cell (see Theorem
\ref{top-R} in Subsection \ref{delicate-part}).
\end{itemize}


The statement (i) follows from good specialization properties of
$M$ and $R$, which are analyzed in Subsection
\ref{specialization}, and Theorem \ref{main-theorem-top}. The
statement (ii) follows from considerations on delineability, and
some properties of real analytic functions. Observe that from (i)
and (ii), a partition of the parameter space (${\Bbb R}^2$, in
this case) such that the topology of $F$ is invariant over each
element of the partition, is computed.


\subsection {Good specialization properties of $M$ and $R$}
\label{specialization}

The aim of this subsection is to prove that the polynomials $M$
and $R$ specialize well out of the curve $R(t,s)=0$, i.e. that
whenever $t-t_0$ (resp. $s-s_0$) is not a factor of $R$, it holds
that $M(x,t_0,s)=M_{t_0}(x,s),R(t_0,s)=R_{t_0}(s)$ (resp.
$M(x,t,s_0)=M_{s_0}(x,t),R(t,s_0)=R_{s_0}(t)$); recall here the
notation $M_{t_i},R_{t_i}$ (resp. $M_{s_i},R_{s_i}$) introduced at
the end of Subsection \ref{hypot}. For this purpose, we begin with
the following result, which can be proven by considering the
Sylvester form of the resultant.

\begin{lemma} \label{tech-result-1}
Let $A_n(t,s)$, $B_m(t,s)$ be the leading coefficients of $F$ and
$M$, respectively, w.r.t. the variables $y$ and $x$, respectively
(note that since by hypothesis $R$ is not identically zero, $F$
depends on $y$, and $M$ depends on $x$). Then the following
statements hold:
\begin{itemize}
\item [(i)] $A_n(t,s)$ is a factor of $M$.
\item [(ii)] $B_m(t,s)$ is a factor of $R$; in particular,
$A_n(t,s)$ is also a factor of $R$.
\end{itemize}
\end{lemma}

This lemma is used for proving the following result.

\begin{lemma} \label{tech-result-2}
Let $t_0\in {\Bbb R}$ satisfy that $t-t_0$ is not a factor of
$R(t,s)$, and let $A_n(t,s)$ be the leading coefficient of $F$
w.r.t. $y$. Then:
\begin{itemize}
\item [(i)] $A_n(t_0,s)\neq 0$; in particular,
$\deg_y(F_{t_0})=\deg_y(F)$.
\item [(ii)] $\Res_y(F,F_y)$ specializes well for $t=t_0$.
\item [(iii)] $F_{t_0}$ (i.e. the specialization $t=t_0$ in $F$)
is square-free as a polynomial in the variables $x,y$.
\end{itemize}
Similarly for $s_0\in {\Bbb R}$, where $s-s_0$ is not a factor of
$R(t,s)$.
\end{lemma}

{\bf Proof.} We prove the statement for $t_0$; similarly for
$s_0$. Now, let us see (i). For this purpose, assume by
contradiction that (i) does not hold. Thus, $A_n(t_0,s)=0$. By the
statement (ii) in Lemma \ref{tech-result-1}, $A_n(t,s)$ is a
factor of $R$; so, $A_n(t_0,s)=0$ implies that $R(t_0,s)=0$, and
therefore $t-t_0$ divides $R$. However, this cannot happen by
hypothesis. Hence (i) follows. Now since (i) holds, we have that
$A_n(t_0,s)\neq 0$, and therefore the statement (ii) follows from
Lemma 4.3.1, pg. 96 in \cite{winkler}. Finally, since (ii) holds,
in case that $F_{t_0}$ is not square-free we have that
$M(x,t_0,s)$ is identically $0$; so, $t-t_0$ divides $M$ and by
Lemma \ref{tech-result-1} it also divides $R$, which cannot happen
by hypothesis. Therefore, (iii) holds. \qed

From the above statement (iii), it may happen that $F_{t_0}$,
where $t-t_0$ is not a factor of $R(t,s)$, has a multiple factor
depending only on $s$, but it cannot have that $F_{t_0}$ has a
multiple factor depending on $x,y,s$. Now in order to see that
$M(x,t_0,s)=M_{t_0}(x,s)$ whenever $t-t_0$ is not a factor of $R$,
we still need an additional property, namely that the
specialization $t=t_0$ of $M(x,t,s)$ has no multiple factor
depending on $x$. This is proven in the following lemma.

\begin{lemma} \label{tech-result-3}
Let $t_0\in {\Bbb R}$ satisfy that $t-t_0$ is not a factor of
$R(t,s)$. Then it holds that:
\begin{itemize}
\item [(i)] $\Res_x(M,M_x)$ specializes well for $t=t_0$.
\item [(ii)] $M(x,t_0,s)$ is square-free as a polynomial in the variable $x$.
\end{itemize}
Similarly for $s_0\in {\Bbb R}$, where $s-s_0$ is not a factor of
$R(t,s)$.
\end{lemma}

{\bf Proof.} Let us see (i). The only case when $\Res_x(M,M_x)$
does not specialize well for $t=t_0$ occurs when the leading
coefficient of $M$ vanishes at $t=t_0$ (see Lemma 4.3.1, pg. 96 in
\cite{winkler}). However, by Lemma \ref{tech-result-1} in that
case $t-t_0$ divides $R$, which cannot happen by hypothesis. So,
(i) holds. Now since (i) holds, if $M(x,t_0,s)$ is not square-free
we have that the specialization of the resultant $\Res_x(M,M_x)$
at $t=t_0$ is identically $0$, and therefore that $R(t_0,s)=0$;
so, $t-t_0$ divides $R$, which cannot happen by hypothesis.
Therefore, (ii) also holds. Similarly for $s=s_0$. \qed

Finally, Lemma \ref{tech-result-2} and Lemma \ref{tech-result-3}
provide the following corollary.

\begin{corollary} \label{corol-R-good-spec}
Let $t_0\in {\Bbb R}$ satisfy that $t-t_0$ is not a factor of
$R(t,s)$. Then the following statements are true:
\begin{itemize}
\item [(i)] $M_{t_0}(x,s)=M(x,t_0,s)$
\item [(ii)] $R_{t_0}(s)=R(t_0,s)$.
\end{itemize}
Similarly for $s_0$, where $s-s_0$ is not a factor of $R$.
\end{corollary}

\subsection{Behavior of the family over the connected components
of ${\Bbb R}^2 \backslash {\mathcal R}$} \label{behav}

Here, we will see that the topology type of the family stays
invariant along each of the connected components of ${\Bbb R}^2
\backslash {\mathcal R}$. For this purpose, the following
proposition is previously required.

\begin{proposition} \label{preserv-top-t}
Let $t_0\in {\Bbb R}$ satisfy that $t_0$ is not a factor of $R$,
and let $s_0,s_1\in {\Bbb R}$, $s_0<s_1$, fulfilling that
$R(t_0,s)$ does not vanish for $s\in [s_0,s_1]$. Then the topology
type of the family $F_{t_0,s}(x,y)$ does not change for $s\in
[s_0,s_1]$.
\end{proposition}

{\bf Proof.} Since by hypothesis $R(t_0,s)$ does not vanish for
$s\in [s_0,s_1]$, then $F_{t_0,s}(x,y)$ does not contain any
factor $s-a$ with $a\in [s_0,s_1]$, i.e. $F_{t_0,s}(x,y)$ does not
identically vanish for $s\in [s_0,s_1]$. Now let $p(s)$ be the
content of $F_{t_0,s}(x,y)$ with respect to $s$, and let
$\hat{F}_{t_0,s}(x,y)$ be the primitive part of $F_{t_0,s}(x,y)$
w.r.t. to $s$. Observe that since $R(t_0,s)$ by hypothesis does
not vanish for $s\in [s_0,s_1]$, then for $s\in [s_0,s_1]$ both
$F_{t_0,s}(x,y)$ and $\hat{F}_{t_0,s}(x,y)$ define the same
family. From the results in Subsection \ref{specialization} it
follows that $\hat{F}_{t_0,s}(x,y)$ fulfills the hypotheses of
Theorem \ref{main-theorem-top}, and that the set of real roots of
$\frac{1}{p(s)}\cdot R(t_0,s)$ is a critical set of the family
$\hat{F}_{t_0,s}(x,y)$. Since $R(t_0,s)$ does not vanish for $s\in
[s_0,s_1]$, the result follows from Theorem
\ref{main-theorem-top}.  \qed

The result in the following proposition is proven in an analogous
way.

\begin{proposition} \label{preserv-top-s}
Let $s_0$ satisfy that $s-s_0$ is not a factor of $R$, and let
$t_0,t_1\in {\Bbb R}$, $t_0<t_1$, fulfilling that $R(t,s_0)$ does
not vanish for $t\in [t_0,t_1]$. Then the topology type of the
family $F_{t,s_0}(x,y)$ does not change for $t\in [t_0,t_1]$.
\end{proposition}

Finally, the following theorem can be proven. Here, the connected
components of ${\Bbb R}^2 \backslash {\mathcal R}$ are denoted as
${\mathcal L}_1,\ldots,{\mathcal L}_p$.

\begin{theorem} \label{top-open-regions}
The topology type of the family $F_{t,s}(x,y)=0$ stays invariant
along each ${\mathcal L}_i$, with $i\in \{1,\ldots,p\}$.
\end{theorem}

{\bf Proof.} Since ${\mathcal L}_i$ is open and connected, one may
always find finitely many segments $L_1,\ldots,L_q$, all of them
lying in ${\mathcal L}_i$, verifying that: (i) each $L_k$ is
either horizontal (in which case $s$ is constant over $L_k$) or
vertical (in which case $t$ is constant over $L_k$); (ii) the
end-point of $L_k$ is the starting-point of $L_{k+1}$; (iii)
$L_1\cup\cdots \cup L_q$ is a path lying in ${\mathcal L}_i$, and
connecting the points $(t_0,s_0)$ and $(t_1,s_1)$. Now by
Proposition \ref{preserv-top-t} and Proposition
\ref{preserv-top-s}, we have that the topology type stays
invariant along each $L_k$. Hence, the result follows. \qed


\subsection{Topology types arising over $R(t,s)=0$}
\label{delicate-part}

If $R$ is not square-free one can always get rid of multiple
factors and keep its square-free part. Thus, in the sequel we
assume that $R$ is square-free. Moreover, we also assume that $R$
is not a univariate polynomial. Observe that if $R=R(t)$
(similarly if $R=R(s)$), then we just have to study the
uniparametric families $F(x,y,a_i,s)$ where the $a_i$'s are the
real zeroes of $R$; this can be done by applying the results in
Subsection \ref{top-one-param}. Now let $t_1,\ldots,t_r$ be the
real zeroes of the discriminant $D_s(R)$ of $R$ w.r.t. to the
variable $s$, where $t_1<\cdots<t_r$, and let ${\mathcal
B}=\{t_1,\ldots,t_r\}$. Then, from \cite{Arnon-Collins} (see pp.
246 in \cite{Arnon-Collins}) one may see that $R$ is delineable
over each interval $(t_k,t_{k+1})$, i.e. that over $(t_k,t_{k+1})$
there exist real analytic functions $\varphi_{k_l}(t)$ (namely,
the real roots of $R$ over $(t_k,t_{k+1})$) verifying that the
graph of $R$ over $(t_k,t_{k+1})$ is the union of the
non-intersecting graphs of the $\varphi_{k_l}(t)$ (see Subsection
\ref{top-one-param} for more information on delineability). In
other words, each $\varphi_{k_l}(t)$ corresponds to a different
analytic branch of $R$ over $(t_k,t_{k+1})$.

Moreover, let \[\tilde{G}(x,y,t)=\sqrt{\Res_s(F,R)}\]One may see
that since by hypothesis $F$ has no factor only depending on
$t,s$, this polynomial is not identically $0$. Nevertheless,
$\tilde{G}$ may contain univariate factors depending on $t$, that
correspond to univariate factors of $R$. These factors
$t-\tilde{t}$ give rise to uniparametric families
$F(x,y,\tilde{t},s)$ that can be analyzed separately by applying
the results in Subsection 2.1; also, notice that the $\tilde{t}$'s
are also real roots of $D_s(R)$. We denote by $G$ the polynomial
obtained by removing the factors $t-\tilde{t}$ from $G$. Since,
from Subsection \ref{hypot}, one may assume that the leading
coefficient of $G$ with respect to $y$ does not depend on $x$, we
have that the polynomial $G(x,y,t)$ verifies the hypotheses of
Theorem \ref{main-theorem-top}. So, let ${\mathcal C}$ be a
critical set of the uniparametric family of parameter $t$ defined
by $G$, and let ${\mathcal A}={\mathcal B}\cup {\mathcal C}$; we
assume that the elements in ${\mathcal A}$ are increasingly
ordered. Moreover, in the sequel we consider an interval $I\subset
{\Bbb R}$ verifying that $I\cap {\mathcal A}=\emptyset$; so, in
particular $R$ is delineable over $I$ and therefore over $I$ the
graph of $R$ is the union of several analytic branches
$\varphi_1(t),\ldots,\varphi_n(t)$. In this situation, the main
result of this subsection is the following.

\begin{theorem} \label{top-R}
Assume that $R$ is delineable over an interval $I$, and let
$\varphi_1(t),\ldots,\varphi_n(t)$ be the real roots of $R$ over
$I$. Then along each $\varphi_j(t)$, with $j\in \{1,\ldots,n\}$,
the topology type of the curves defined by
$F(x,y,t,\varphi_j(t))$, with $t\in I$, stays invariant.
\end{theorem}

In other words, the theorem states that, whenever one moves along
an analytic branch of $R$ in between two consecutive elements of
${\mathcal A}$, the topology type of the family is preserved.
Thus, a finite partition of $R(t,s)=0$ into 1-dimensional and
0-dimensional cells can be computed so that the topology type of
the family remains invariant along each element of the partition.
The result is illustrated in Figure 2; in this picture, ${\mathcal
A}=\{t_1,t_2,t_3,t_4,t_5\}$. The rest of the subsection is devoted
to proving this result.

\begin{figure}[ht]
\begin{center}
\centerline{
\psfig{figure=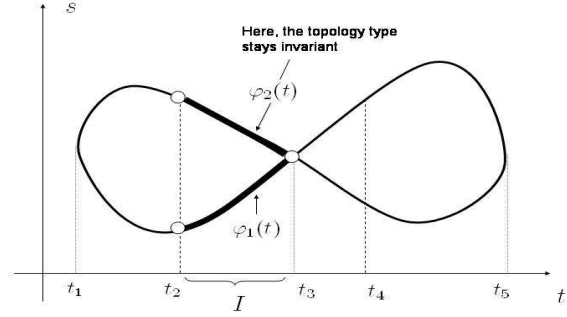,width=8cm,height=5cm}}
\end{center}
\caption{Illustrating Theorem \ref{top-R}}
\end{figure}

In order to prove the theorem, some previous results are needed.
The first result states that the zero-set of $R(t,s)$ over $I$ can
be expressed as the union of certain analytic functions.

\begin{lemma} \label{analytic-R}
Let $R(t,s)=\alpha_p(t)s^p+\alpha_{p-1}(t)s^{p-1}+\cdots$. Then,
there exist $p$ different analytic functions
$\psi_1(t),\ldots,\psi_p(t)$ so that the zero set of $R(t,s)$ over
$I$ is the union of the zero-sets of the functions $s=\psi_k(t)$,
$k=1,\ldots,p$.
\end{lemma}

{\bf Proof.} $n$ of these functions are the real roots of $R$ over
$I$. The existence of the remaining $p-n$ (complex) functions
follows, for instance, from the complex version of the Implicit
Function Theorem (see p. 84 in \cite{Abhyankar}) and analytic
continuation. \qed

In fact, from the complex version of the Implicit Function Theorem
it follows that the $\psi_k(t)$'s are defined over open complex
subsets (containing $I$); so, $R(t,s)$ is defined over an open
complex subset $U$ whose $t$-projection $\pi_t(U)$ contains $I$.
Now Lemma \ref{analytic-R} is required for proving the following
result.

\begin{lemma} \label{zero-set-G}
The zero set of $G(x,y,t)$ over $I$ (i.e. the zero set of $G$
whose $t$-projection is $I$) is the union of the
zero-sets over $I$ of the functions
$F_j(x,y,t)=F(x,y,t,\psi_j(t))$ (in the sequel, $F_j$),
$j=1,\ldots,p$.
\end{lemma}

{\bf Proof. }By definition the zero set of $G(x,y,t)$ over $I$ is
the zero set over $I$ of $H(x,y,t)$, where $H$ is the result of
removing from $\Res_s(F,R)$ the univariate factors depending on
$t$. From Lemma \ref{analytic-R} and properties of the resultant
(see property 3, page 255 in \cite{sendra-winkler}) we have that
$\Res_s(F,R)=\alpha_p(t)\cdot F_1(x,y,t)\cdots F_p(x,y,t)$. Then
removing the univariate factors corresponding to $\alpha_p(t)$ we
get that zero set of $H$ coincides with that of $F_1\cdots F_p$.
\qed

Let $U^{\star}$ be the projection onto $(x,y,t)$ of the open
subset ${\Bbb R}^2\times U$. Then, one may see that for
$j=1,\ldots,p$, $F_j$ is analytic over $U^{\star}$ (because it is
the composition of two analytic functions, namely $F$ and
$\psi_j(t)$), and writing $\bar{F}=F_1\cdots F_p$, so is
$\bar{F}$; notice that $\pi_t(U^{\star})$ contains
$I$. Hence, for each point $P\in U^{\star}$ there exists an open
(complex) subset $U_p\subset U^{\star}$ so that $\bar{F}$ has
can be expanded as a power series convergent in $U_p$; in this 
situation, we
say that $\bar{F}$ defines
a {\it germ} over $U_p$ (i.e. the zero set of $\bar{F}$ over $U_p$, see
\cite{Jong} or \cite{Abhyankar} for further information on germs).
Moreover, because of Lemma \ref{zero-set-G}, $G$ defines
the same germ. Now the following lemma is the key for proving
Theorem \ref{top-R}. Here we will use some ideas and results from
Analytic Geometry related to germs. Namely, we will use the notion
of irreducible germ, and the fact that every germ can be uniquely
written as an ``irredundant" union of irreducible germs (i.e. as a
finite union of all-distinct, irreducible germs). We refer to
Chapter V of \cite{Abhyankar} and Chapters 3, 4 of \cite{Jong} for
further reading on these questions. Also, we will use the
following notation, analogous to the notation in Lemma
\ref{real-roots} (see Subsection 2.2). Here we have tried to
simplify the description in order to avoid a cumbersome notation.

\begin{itemize}
\item The real roots of the square-free part of the discriminant
$D_y(G)$ over $I$ are denoted as $X_k$'s (observe that since $I$
contains no point of the critical set of $G$, from Lemma
\ref{real-roots} the square-free part of $D_y(G)$ is delineable
over $I$); the graph of $X_k$ is denoted as ${\mathcal X}_k$.
\item From Lemma \ref{real-roots}, $G$ is delineable over each ${\mathcal
X}_k$. The real roots of $G$ over ${\mathcal X}_k$ are denoted as
$Y_l$'s; the graph of $Y_l$ is denoted as ${\mathcal Y}_l$.
\item We denote by $C_k$ the region of the
$xt$-plane lying in between two consecutive ${\mathcal X}_k$'s,
with $t\in I$. Also from Lemma \ref{real-roots}, we have that $G$
is delineable over $C_k$. The real roots of $G$ over $C_k$ are
denoted as $V_i$'s; the graph of $V_i$ is denoted as ${\mathcal
V}_i$.
\end{itemize}

Hence, the following lemma holds. Essentially, this result ensures
that each ${\mathcal V}_i$ is associated with some $\psi_j(t)$'s,
and conversely. The first statement of this lemma is illustrated
in Figure 3.

\begin{lemma} \label{key-lemma}
The following statements hold:
\begin{itemize}
\item [(a)] Let ${\mathcal V}_i$ be a real root of $G$ over a region $C_k$.
Then there exists $j\in \{1,\ldots,p\}$, satisfying that
$F(x,y,t,\psi_j(t))|_{{\mathcal V}_i}=0$.
\item [(b)] Let $j\in \{1,\ldots,p\}$, and let $t_0\in I$. If
$F(x,y,t_0,\psi_j(t_0))$ vanishes at ${\mathcal V}_i \cap
\{t=t_0\}$, then $F(x,y,t,\psi_j(t))|_{{\mathcal V}_i}=0$.
\end{itemize}
\end{lemma}

{\bf Proof.} Let us see first the statement (a). From Lemma
\ref{zero-set-G}, it follows that ${\mathcal V}_i$ is included in
the zero-set of $\bar{F}=F_1\cdots F_p$. Moreover, each $F_j$ is
analytic over $U^{\star}$, and ${\mathcal V}_i \subset U^{\star}$.
So, for each point $P\in {\mathcal V}_i$ there exists an open
complex subset $U_p\subset U^{\star}$ containing $P$ so that each
$F_j$, and therefore also $\bar{F}$, defines a germ over $U_p$; in
the rest of the proof we will refer to these germs as the ``zero
sets" of $F_j$, $\bar{F}$ over $U_p$, respectively. Furthermore, since by
Lemma \ref{vis-connected} ${\mathcal V}_i$ is connected, we can
always take $U_p$ sufficiently small so that ${\mathcal V}_i\cap
U_p$ is also connected. Now, the zero-set of each $F_j$ over $U_p$ can be
written as a finite union of irreducible germs (see p. 237 of
\cite{Abhyankar}) $W_j^1,\ldots,W_j^{l_j}$. Moreover, the zero-set
of $\bar{F}$ over $U_p$ can also be written as an ``irredundant" union of
irreducible germs $\tilde{W}_1\cup \cdots \cup \tilde{W}_q$, and
each $\tilde{W}_r$ is included in some $W_a^b$, where $a\in
\{1,\ldots,p\}$ and $b\in \{1,\ldots, l_a\}$ (see p. 240 of
\cite{Abhyankar}). Let us see that there exists just one $s\in
\{1,\ldots,q\}$ verifying that ${\mathcal V}_i \cap U_p \subset
\tilde{W}_s$. Indeed, clearly ${\mathcal V}_i \cap U_p \subset
\tilde{W}_1\cup \cdots \cup \tilde{W}_q$. Now if the statement
does not hold then either ${\mathcal V}_i\cap U_p$ is not
connected, which cannot happen, or there exist two different
$\tilde{W}_r$'s, say $\tilde{W}_A, \tilde{W}_B$, and a point $Q\in
{\mathcal V}_i\cap U_p$, so that $Q\in \tilde{W}_A\cap \tilde{W}_B$.
However, since $\tilde{W}_A, \tilde{W}_B$ are different germs
in this last case $Q$ would
be a
self-intersection of the surface $S_G$ defined by $G$, and therefore a singular point of $S_G$;
but this cannot happen, either,
because from Corollary \ref{singular-surface} every singular point
of $S_G$ with $t\in I$ projects onto some ${\mathcal X}_k$. So,
there exists $s\in \{1,\ldots,q\}$ so that ${\mathcal V}_i \cap
U_p \subset \tilde{W}_s$. Then, let $j\in \{1,\ldots,p\}$ satisfy
that $\tilde{W}_s\subset W_j^b$, where $b\in \{1,\ldots,l_j\}$.
Hence, $F(x,y,t,\psi_j(t))|_{{\mathcal V}_i\cap U_p}=0$. Finally,
since $F(x,y,t,\psi_j(t))$ is analytic, is defined over the whole
${\mathcal V}_i$, and vanishes over ${\mathcal V}_i \cap U_p$,
then it vanishes over the whole ${\mathcal V}_i$ (see p. 81 in
\cite{Jong}).

In order to prove part (b), by contradiction one assumes that the
statement is not true, and, reasoning as in part (a), one shows that the surface $S_G$ has a self-intersection
not projecting onto any ${\mathcal X}_k$, which violates Corollary \ref{singular-surface}.

 \qed

\begin{figure}[ht]
\begin{center}
\centerline{
\psfig{figure=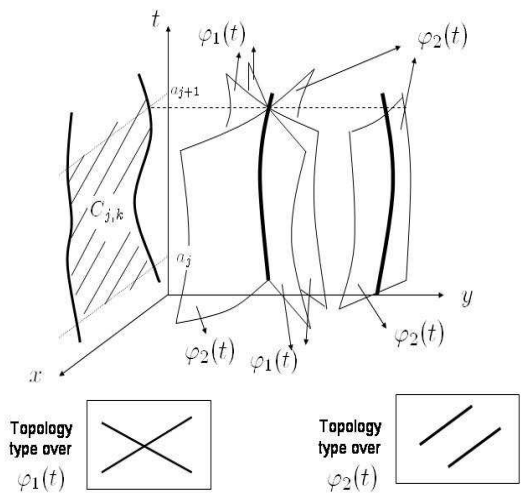,width=8cm,height=7cm}}
\end{center}
\caption{Illustrating part (a) of Lemma \ref{key-lemma}}
\end{figure}

\begin{remark} Notice that the $j$ appearing in the statement (a) of Lemma
\ref{key-lemma} is not necessarily unique, i.e. it may happen that
given ${\mathcal V}_i$ there exist $j_1\ldots,j_{m_i}$ so that
$F(x,y,t,\psi_{j_1}(t))|_{{\mathcal
V}_i}=\cdots=F(x,y,t,\psi_{j_{m_i}}(t))|_{{\mathcal V}_i}=0$. Moreover,
part (b) of Lemma
\ref{key-lemma} essentially says that over a $C_k$, two different $F_j$'s
are either disjunct or fully coincident;
therefore, a $F_j$ (i.e. its zero-set) cannot contain a part of a ${\mathcal V}_i$, but a whole ${\mathcal V}_i$.
\end{remark}

Finally, Theorem \ref{top-R} can be proven.

{\bf Proof of Theorem \ref{top-R}.} Let $\varphi_j(t)$, with $j\in
\{1,\ldots,n\}$, be a real root of $R$ over $I$. Now from Lemma \ref{zero-set-G}, the
zero-set of $F_j(x,y,t)=F(x,y,t,\varphi_j(t))$ with $t\in I$ is included in the zero-set of $G$.
Moreover, for each region $C_k$ the zero-set of $G$ over $C_k$ is equal to the union of
the ${\mathcal V}_i$'s; then, from Lemma
\ref{key-lemma}, for each region $C_k$ there exists a subset
$J_k=\{k_1,\ldots,k_a\}$ so that the real part of
the zero-set of $F(x,y,t,\varphi_j(t))$ with $t\in I$ and $(x,t)\in C_k$
 is equal to
the union of the ${\mathcal V}_i$'s with $i\in J_k$ ($J_k$ is empty iff $F_j(x,y,t)$ has no
real zero with $t\in I$ and $(x,t)\in C_k$). Furthermore, if
$F(x,y,t,\varphi_j(t))|_{{\mathcal V}_i}=0$ and ${\mathcal Y}_l$
is in the closure of ${\mathcal V}_i$, then
$F(x,y,t,\varphi_j(t))|_{{\mathcal Y}_l}=0$ because $F_j$ is continuous. Finally, since by
Lemma \ref{main-point} the relative positions of the ${\mathcal
V}_i$'s, the ${\mathcal Y}_l$'s, and of the ${\mathcal V}_i$'s
w.r.t. the ${\mathcal Y}_l$'s stay invariant when $t\in I$, we have that
the topology type of the level curves of $F_j(x,y,t)=0$ with $t\in I$
stays invariant. Hence, Theorem \ref{top-R} follows. \qed

\subsection{The Algorithm} \label{algorithm}

From the ideas in the preceding subsections, we can derive the
following algorithm for computing a finite partition of ${\Bbb
R}^2$ into 0-dimensional, 1-dimensional and 2-dimensional cells so
that the topology type of the family defined by $F$ stays
invariant along each cell; we denote by $C_{[0]},C_{[1]},C_{[2]}$
the sets consisting of all the 0-dimensional cells, the
1-dimensional cells, and the 2-dimensional cells, respectively.
Here, we assume that $F$ fulfills the hypotheses made explicit at
the beginning of the section, and that $R\neq 0$. Observe that
once the partition has been computed, the topology types in the
family might be determined by first choosing a point $(t_i,s_i)$
in each partition element, and then applying the method in
\cite{Lalo}, \cite{Hong} for describing the topology of the
resulting curve. However, in some cases it can be difficult or
even impossible to choose $t_i,s_i$ both being rational; so, in
some situations we might not obtain all the topology types in the
family. Still, however, we get the parameter values corresponding
to each type.

{\bf \underline{Algorithm:} (two-parameters case)}

\begin{itemize}
\item [1.] [Polynomials $R,\tilde{G},G$] Compute the polynomials $R,\tilde{G},G$.
\item [2.] [Set ${\mathcal A}$] Compute the real roots of
$D_s(R)$, $D_x(\sqrt{D_y(G)})$, and let ${\mathcal A}$ be the set
consisting of these values. Let $I_1,\ldots,I_m$ be the real
intervals verifying that ${\Bbb R}-{\mathcal A}=I_1\cup \cdots
\cup I_m$.
\item [3.] [0-dimensional cells] For all $t_i\in {\mathcal A}$,
where $t-t_i$ does not divide $R$, compute the points
$P_{1,i},\ldots,P_{r_i,i}$ verifying that $R(t_i,s)=0$. Then
\[C_{[0]}=\bigcup_i \bigcup_{m=1}^{r_i}\{P_{m,i}\}\] Some other points may be
added in Step 4.1.
\item [4.] [1-dimensional cells]
\begin{itemize}
\item [4.1] [Univariate factors] For each $\tilde{t}_i$ where
$t-\tilde{t}_i$ divides $R$, compute a critical set ${\mathcal
P}_i$ of the family defined by $F(x,y,\tilde{t}_i,s)$. Let
$J_{1,i},\ldots,J_{n_i,i}$ be the real intervals verifying that
${\Bbb R}-{\mathcal P}_i=J_{1,i}\cup \cdots \cup J_{n_i,i}$, and
let $Q_{k,i}=\tilde{t}_i\times J_{k,i}$. Moreover, add the points
$(\tilde{t}_i,s)$, where $s\in {\mathcal P}_i$, to the list
$C_{[0]}$ of 0-dimensional cells computed in Step 3.
\item [4.2] [Analytic branches of $R$] Let $\varphi_{1,j},\ldots,\varphi_{k_j,j}$ be the
real roots of $R$ over each $I_j$, and let $A_{l,j}=\{(t,s)\in
{\Bbb R}^2|t\in I_j,s=\varphi_l(t)\}$.
\item [4.3] [List of Cells] \[C_{[1]}=\left(\bigcup_{i,k}\{Q_{k,i}\}\right)\cup
\left(\bigcup_{l,j}A_{l,j}\right)\]
\end{itemize}
\item [5.] [2-dimensional cells] Let $B_{i,j}=\{(t,s)\in {\Bbb
R}^2|t\in I_j, \varphi_i(t)<s<\varphi_{i+1}(t)\}$, where
$\varphi_i, \varphi_{i+1}$ denote consecutive real roots of $R$
over $I_j$. Then \[C_{[2]}=\bigcup_{i,j}B_{i,j}\]
\end{itemize}

Observe that, from Theorem \ref{top-open-regions}, if two adjacent
2-dimensional cells computed in the step (5) of the above
algorithm are not separated by any 1-dimensional cell computed in
the step (4), then the topology type of the family is the same
over both cells; in fact, in that case both cells would correspond
to the same connected component of ${\Bbb R}^2 \backslash
{\mathcal R}$. Notice also that two adjacent cells might give rise
to the same topology type; so, the decomposition computed by the
above algorithm is not necessarily minimal. Finally, observe also
that, as in the one-parameter case, whenever $D_y(F)$ is
square-free the ideas of \cite{AMS-Mourrain} and
\cite{MEGA-Lazard} might be used in order to more efficiently
compute $R$.

\subsection{The special case $R=0$}\label{sec-special}

If $R=0$, from the definition of $R$ it holds that either $M=0$,
in which case $F=F(x,t,s)$, or $\deg_x(M)=0$, in which case
$M=M(t,s)$. In both situations the reasonings are completely
analogous to the case $R \neq 0$; so, here we state the main
results for this special case and we leave the proofs to the
reader.

If $M=0$, we denote $P=P(t,s)=\sqrt{D_x(F)}$, which defines a
curve ${\mathcal P}$. We denote the connected components of ${\Bbb
R}^2\backslash{\mathcal P}$ as $A_i$. Moreover, we also denote
$J(x,t)=\Res_s(P,F)$. Hence, $J$ defines a uniparametric family,
and therefore one may compute a critical set ${\mathcal J}$ of the
family. Then, the following result holds. Here, ${\mathcal E}$
denotes the union of the real roots of $D_s(P)$ and the elements
of ${\mathcal J}$.

\begin{theorem} \label{theorem-sec-special}
The topology type of the family $F$ stays the same over each
$A_i$, and also along each real root of $P$ over each interval of
${\Bbb R}$ lying in between two consecutive elements of ${\mathcal
E}$.
\end{theorem}

For the case $\deg_x(M)=0$, $M=M(t,s)$ defines a curve ${\mathcal
M}$; we represent the connected components of ${\Bbb R}^2
\backslash {\mathcal M}$ as $B_j$. Moreover, we write
$K(x,y,t)=\Res_s(F,M)$, and we denote by ${\mathcal K}$ a critical
set of the uniparametric family defined by $K$. Also, ${\mathcal
F}$ denotes the union of the real roots of $D_s(M)$, and the
elements of ${\mathcal F}$. Then we have the following theorem.

\begin{theorem} \label{second-theorem-sec-special}
The topology type of the family $F$ stays the same over each
$B_j$, and also along each real root of $M$ over each interval of
${\Bbb R}$ lying in between two consecutive elements of ${\mathcal
F}$.
\end{theorem}

\section{Examples.}
\label{sec-generalization}

In this section we provide three examples in order to illustrate
the ideas of Section \ref{general-case}. The two first ones
correspond to the case $R\neq 0$, while the third one corresponds
to $R=0$. Moreover, in the second example the topology types
arising in the {\it offset} family to the parabola $y^2-2px=0$ are
computed. Offset curves (see for example \cite{Rafa} for more
information on this subject), widely used in the CAGD context, can
be intuitively described as ``parallel" curves to a given curve at
a certain distance. If the offsetting distance $d$ is not
particularized, then the offset family to a given algebraic curve
is certainly a family of algebraic curves depending on the
parameter $d$, and the topology types in the family can be
computed by using the results in Subsection \ref{top-one-param}
(see \cite{JGRS} for further details); this may be useful in order
to identify the distances where the topology of the offset
coincides with that of the original curve, which is the desired
situation in most applications. Now if the original curve depends
on one parameter, as it happens in the case of $y^2-2px=0$, then
the offset is a family of algebraic curves depending on two
parameters, and therefore the results in our paper are applicable.
The computation of the topology types in the offset family to
$y^2-2px=0$ was solved by Prof. W. L\"u (1992) by using ``ad-hoc"
methods. However, here we compute them as a direct application of
the general algorithm provided in Section \ref{general-case}.

\begin{example} \label{Cassini}
Consider the family of algebraic curves defined by
\[F(x,y,t,s)=(x^2+y^2+t^2)^2-4t^2x^2-s^4=0.\]The curves of this
family are usually known as the {\it Cassini's ovals}. Let us see
how the algorithm works in this case:

\begin{itemize}
\item [1.] [$R,\tilde{G},G$] The polynomial $R$ (after removing multiple factors) is
\[R(t,s)=st(2t^2-s^2)(2t^2+s^2)(-s+t)(t+s)(t^2+s^2)\]Moreover,
we also get

$\tilde{G}(x,y,t)=(x^4+2x^2y^2-2t^2x^2+y^4+2y^2t^2+t^4)t\cdot \\
(x^4+2x^2y^2-2t^2x^2+y^4+2y^2t^2-3t^4) \cdot\\
(x^4+2x^2y^2-2t^2x^2+y^4+2y^2t^2)$

One may see that $\tilde{G}$ has just one univariate factor,
namely $t$, depending on the variable $t$. Then $G$ is immediately
obtained.
\item [2.] [${\mathcal A}$] $D_s(R)$, $D_x(\sqrt{D_y(G)})$ have just one real root,
namely $0$; so, ${\mathcal A}=\{0\}$.
\item [3.] [0-dimensional cells] We have just one 0-dimensional cell, namely
$\{(0,0)\}$.
\item [4.] [1-dimensional cells]
\begin{itemize}
\item [4.1] [Univariate factors] Over $t=0$, the family reduces to $(x^2+y^2)^2-s^4=0$.
A critical set of this new family is $\{0\}$. So, no new
0-dimensional cells are found, and we get two 1-dimensional cells,
namely $\{0\}\times (0,\infty)$ and $\{0\}\times (-\infty,0)$.
\item [4.2] [Analytic branches of $R$] When $t\in (-\infty,0)$, $R$ has 5 real
roots, corresponding to the cases $s=0$, $s=\sqrt{2}t$,
$s=-\sqrt{2}t$, $s=t$, $s=-t$, respectively (see Figure 4); each
one gives rise to a different 1-dimensional cell. The same happens
when $t\in (0,\infty)$.
\end{itemize}
\item [5.] [2-dimensional cells] They are the two-dimensional
regions lying in between consecutive real roots of $R$ over
$(-\infty,0)$ and $(0,\infty)$, respectively (see also Figure 4).
One may see in Figure 4 that there are 12 of these cells, named as
$I, II, \ldots, XII$; also, the border between, say, $I$ and $II$
corresponds to the 1-dimensional cell defined by $t\in (0,\infty)$
and $s=\sqrt{2}t$, etc.
\end{itemize}

One may find the topology types corresponding to each cell also in
Figure 4.

\begin{figure}[ht]
\begin{center}
\centerline{ \psfig{figure=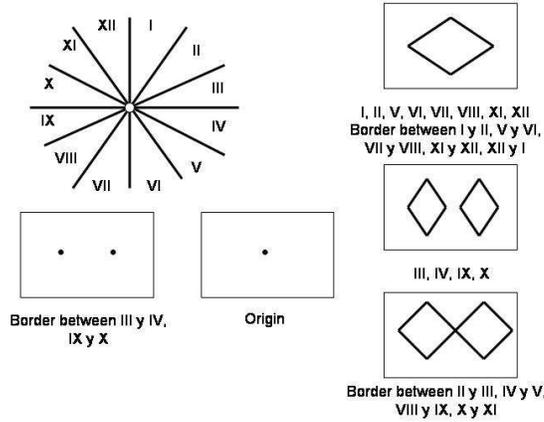,width=8cm,height=6cm}}
\end{center}
\caption{Topology types in Example 1}
\end{figure}

\end{example}

\begin{example}
Consider the family of parabolas defined by $y^2-2px=0$. One may
check that the equation of the corresponding offset family is

$F(x,y,p,d)=-8d^2y^2x^2+y^4p^2+4x^2y^4+4y^6-12d^2y^4+12y^2d^4+4d^4x^2
-4d^6-20p^2d^2y^2+4py^2xd^2-4p^4d^2-8p^2d^4-8p^2d^2x^2-16px^3d^2
-16px^3y^2+32p^2y^2x^2-4p^3y^2x-20pxy^4+16pxd^4+16p^3d^2x+16p^2x^4-16p^3x^3+4p^4x^2$

Moreover, the computation of the double discriminant yields (after
removing multiple factors):
\[R(p,d)=pd(p+8d)(p-d)(p-8d)(p+d)(8p^2+d^2)\]
Without loss of generality we can assume that $p\neq 0$ (otherwise
a degenerated situation is reached) and $d\neq 0$ (the offsetting
distance is never $0$); moreover, also w.l.o.g. we can assume that
$p>0$, $d>0$. One can check that in this case a critical set of
the polynomial $\tilde{G}$ reduces to $\{0\}$; so, we have the
following cases: (1) $0<p<d$; (2) $p=d$; (3) $d<p<8d$; (4) $p=8d$;
(5) $p>8d$. Furthermore, one may also check that the topology type
coincides in (3), (4) and (5); so, finally we get three topology
types corresponding to the cases $p<d$, $p=d$, $p>d$,
respectively, which are shown in Figure 5.

\begin{figure}[ht]
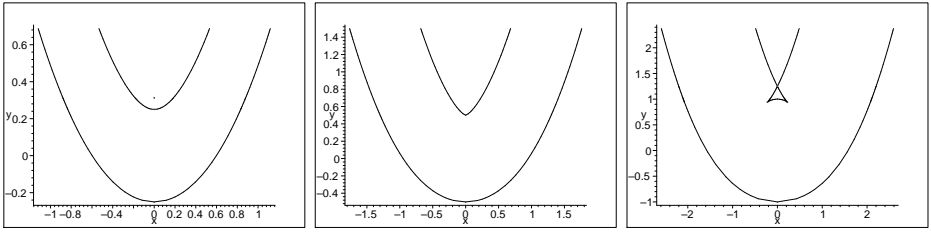

\begin{center}
\centerline{$\begin{array}{ccc}
\psfig{figure=Offset_parabola01.eps,width=4cm,height=3cm} &
\psfig{figure=Offset_parabola02.eps,width=4cm,height=3cm} &
\psfig{figure=Offset_parabola03.eps,width=4cm,height=3cm}
\end{array}$}
\end{center}

\caption{Topology types of the offsets to $y^2-2px=0$: $p<d$
(left); $p=d$ (center); $p>d$ (right)}
\end{figure}

\end{example}

\begin{example} \label{ex-esp-case}
We consider the linear system of curves defined by
\[F(x,y,t,s)=-1+x^2+t(x-y)+s(x^3-y)\]Here, we get that
$M(x,t,s)=-t-s$. Thus, $R(t,s)=0$. Then we consider the
uniparametric family defined by
$K(x,y,t)=\Res_s(F,M)=-1+x^2+tx-tx^3$. Since $K$ does not depend
on $y$, from Theorem \ref{main-theorem-top} we have that the set
of real roots of $D_x(K)=4t^5-8t^3+4t$, i.e. $\{-1,0,1\}$, is a
critical set of the family. Hence, these values induce a partition
of the line $-t-s=0$ into four pieces, corresponding to the cases
$t\in (-\infty,-1)$, $t\in (-1,0)$, $t\in (0,1)$ and $t\in
(1,\infty)$, respectively. More precisely, we have the following
partition of the parameter space:

\begin{itemize}
\item $\mbox{[0-dimensional cells]}$: $\{(-1,1)\},\{(0,0)\},\{(1,-1)\}$; here, the
topology type of $F$ is that of two parallel lines (in the three
cases).
\item $\mbox{[1-dimensional cells]}$: $\{(t,s)\in {\Bbb R}^2|t\in
(-\infty,-1),-t-s=0\}$; $\{(t,s)\in {\Bbb R}^2|t\in
(-1,0),-t-s=0\}$; $\{(t,s)\in {\Bbb R}^2|t\in (0,1),-t-s=0\}$;
$\{(t,s)\in {\Bbb R}^2|t\in (1,\infty),-t-s=0\}$; here, the
topology type is that of three parallel lines (in all the cases).
\item $\mbox{[2-dimensional cells]}$: $\{(t,s)\in {\Bbb
R}^2|-t-s>0\}$; $\{(t,s)\in {\Bbb R}^2|-t-s<0\}$; here, the
topology type is that of a line (in both cases).
\end{itemize}
\end{example}

\end{document}